\documentclass[12pt]{article}

\begin{document}
\setcounter{page}{1}
\renewcommand{\thefootnote}{\fnsymbol{footnote}}
\pagestyle{plain} \vspace{1cm}
\begin{center}
\large{\bf Entropic force approach to noncommutative Schwarzschild
black holes signals a failure of current physical ideas}

\small \vspace{1cm} {\bf S. Hamid
Mehdipour\footnote{mehdipour@liau.ac.ir}}
\\
\vspace{0.5cm} {\it Islamic Azad University, Lahijan Branch,
P. O. Box 1616, Lahijan, Iran }\\

\end{center}
\vspace{1.5cm}

\begin{abstract}
Recently, a new perspective of gravitational-thermodynamic duality
as an entropic force arising from alterations in the information
connected to the positions of material bodies is found. In this
paper, we generalize some aspects of this model in the presence of
noncommutative Schwarzschild black hole by applying the method of
coordinate coherent states describing smeared structures. We
implement two different distributions: (a) Gaussian and (b)
Lorentzian. Both mass distributions prepare the similar quantitative
aspects for the entropic force. Our study shows, the entropic force
on the smallest fundamental unit of a holographic screen with radius
$r_0$ vanishes. As a result, black hole remnants are unconditionally
inert even gravitational interactions do not exist therein. So, a
distinction between gravitational and inertial mass in the size of
black hole remnant is observed, i.e. the failure of the principle of
equivalence. In addition, if one considers the screen radius to be
less than the radius of the smallest holographic surface at the
Planckian regime, then one encounters some unusual dynamical
features leading to gravitational repulsive force and negative
energy. On the other hand, the significant distinction between the
two distributions is conceived to occur around $r_0$, and that is
worth of mentioning: at this regime either our analysis is not the
proper one, or non-extensive
statistics should be employed.
\end{abstract}
\newpage

\section{\label{sec:1}Introduction}
The exploration of thermal radiation from black holes has mainly
emerged a close relationship between black hole physics and
thermodynamics \cite{haw}. One striking clue for the nature of
gravity comes from the profound investigation of black hole
thermodynamics due to the fact that it is possible to provide a
physical resemblance between spacetimes including horizons and the
notions of temperature or entropy. In addition, a quantum theory of
gravity tells us that the black hole entropy could be related to a
number of microscopic states. Therefore, detailed studies of the
black hole entropy may have important implications for a complete
theory of quantum gravity. This is the main reason why the origin of
the black hole entropy requires to be perceived at the fundamental
level. In 1995, Jacobson exhibited that the Einstein equations are
acquired from the laws of thermodynamics \cite{jac}. Padmanabhan
also used the argument of equipartition energy of horizons to
prepare a thermodynamic perspective of gravity \cite{pad1}.
Recently, Verlinde has proposed an emergent phenomenon for the
origin of Newtonian gravity and general relativity \cite{ver}. This
theory implies that gravitational interaction arises from the
statistical behavior of microscopic degrees of freedom encoded on a
holographic screen and can be described as a kind of entropic force,
related to the information that is stored on the holographic
surfaces. The idea of entropic force in different cases has been
investigated by many authors \cite{ent}. Also, there are some
comments on the entropic gravity scenario which indicate some short
comings of this scenario in addition to its open problems
\cite{comm}, they can appear as the topic of a new debate.

If the gravitational force is entropic and entropy couples the
emergent picture of gravity with the fundamental microstructure of a
quantum spacetime, then in Verlinde's formalism we should
investigate the microscopic scale effects by using tools such as
noncommutative gravity for the interpretation of the microscopic
structure of a quantum spacetime. Perhaps one way of observing
noncommutativity is through assessing its influences on the
properties of black holes. Nicolini {\it et al} \cite{nic1} in a new
conceptual approach to noncommutative gravity, based on coordinate
coherent state formalism, have improved the short distance behavior
of point-like structures. In their method, curvature singularities
which appear in general relativity, can be eliminated. They have
demonstrated that black hole evaporation process should be stopped
when a black hole reaches a minimal mass. This minimal mass, named
{\it black hole remnant}, is a result of the existence of a minimal
observable length. This approach, which is the so-called
{\it{noncommutative geometry inspired model}}, via a minimal length
caused by averaging noncommutative coordinate fluctuations
\cite{sma} cures the curvature singularity in black holes. In fact,
the curvature singularity at the origin of black holes is
substituted for a regular deSitter core. Accordingly, the ultimate
phase of the Hawking evaporation as a novel thermodynamically steady
state comprising a non-singular behaviour is concluded (for more
details, see \cite{nic2}). It must be noted that, generally, it is
not required to consider the length scale of the coordinate
noncommutativity to be the same as the Planck length. Since, the
noncommutativity influences appears on a length scale connected to
that region, they can behave as an adjustable parameter
corresponding to that pertinent scale.

In this paper, we use noncommutative geometry inspired model to
combine the microscopic structure of spacetime with the entropic
description of gravity because the concept of entropy has a profound
relationship with the quantum spacetime structures. The layout of
the paper is as follows. We begin in Sec.~\ref{sec:2} by outlining
the entropic force approach. In Sec.~\ref{sec:3}, the entropic force
in the presence of noncommutative Schwarzschild black hole is
derived by considering the effect of smearing of the particle mass
as Gaussian and Lorentzian distributions. Finally, a summary and the
conclusion follows in Sec.~\ref{sec:4}.

\section{\label{sec:2}Entropic Force Approach}
In this section we briefly investigate Verlinde's approach.
Afterwards, by considering the case of noncommutative geometry
inspired Schwarzschild black hole, we try to improve the expression
of black hole's entropic force by taking into account the
noncommutative corrections at the scale which noncommutativity
influences set in. In order to obtain the temperature from an
entropic force in general relativity, we first consider a static
background with a global timelike Killing vector $\xi^\mu$. To
define a foliation of space, and viewing the holographic screens
$\Omega$ as surfaces of constant redshift, we write the potential
$\phi$ as \cite{ver} {\footnote{In this work natural units are used
with the following definitions: $\hbar = c = G = k_B = 1$.}}
\begin{equation}
\label{mat:1}\phi=\frac{1}{2}\log\left(-\xi^\mu\xi_\mu\right),
\end{equation}
where $\mu=0,\,1,\,2,\,3$. Note that $e^\phi$ indicates the redshift
factor and it makes a connection between the local time coordinate
and the reference point with $\phi = 0$ at infinity.

The four acceleration $a^\mu$, for a particle that is placed very
adjacent to the screen can be written as
\begin{equation}
\label{mat:2}a^\mu=-\nabla^\mu\phi.
\end{equation}
As can be seen, the acceleration is perpendicular to the holographic
screen. Then, with defining the normal vector
$N^\mu=\frac{\nabla^\mu\phi}{\sqrt{\nabla^\nu\phi\nabla_\nu\phi}}$,
the local temperature on the screen is achieved by
\begin{equation}
\label{mat:3}T=-\frac{1}{2\pi}e^\phi N^\mu
a_\mu=\frac{1}{2\pi}e^\phi\sqrt{\nabla^\mu\phi\nabla_\mu\phi}.
\end{equation}
The change in entropy for a test particle with mass $m$ at fixed
position close to the screen equals
\begin{equation}
\label{mat:4}\nabla_\mu S=-2\pi m N_\mu.
\end{equation}
The entropic force is now found to have the form
\begin{equation}
\label{mat:5}F_\mu=T\nabla_\mu S=-me^\phi\nabla_\mu\phi.
\end{equation}
The above expression is indeed the relativistic analogue of the
second law of Newton $F=ma$. The redshift factor $e^\phi$ is added
due to the fact that the gravitational force is measured with
respect to the reference point at infinity.

Let us now suppose that the energy $E$ associated with the mass $M$
(assumedly larger than the test particle of mass $m$ and is located
at the origin of the coordinate as the source) is distributed on a
closed surface of constant redshift $\phi$. On this screen, $N$ bits
of information are stored and the holographic information from mass
$M$ is encoded as $dN=dA$ \cite{pad2}, where $A$ is the area of the
screen. The energy on the screen obeys thermal equipartition,
\begin{equation}
\label{mat:6}E=\frac{1}{2}\int_\Omega TdN=\frac{1}{4\pi}\int_\Omega
e^\phi\nabla\phi dA.
\end{equation}
This result is in agreement with the Gauss's theorem.

\section{\label{sec:3}Noncommutative Geometry Inspired Model: Gaussian and Lorentzian Distributions}
We now consider the metric of noncommutative Schwarzschild black
hole. According to Refs.~\cite{nic1,nic2}, the mass density of a
static, asymptotically flat, spherically symmetric, particle-like
gravitational source cannot be a delta function distribution but
will be given by a Gaussian distribution of minimal width
$\sqrt{\theta}$ as follows
\begin{equation}
\label{mat:7}\rho_{\theta}(r)=\frac{M}{(4\pi\theta)^{\frac{3}{2}}}e^{-\frac{r^2}{4\theta}}.
\end{equation}
The metric of the Einstein equations connected to these smeared mass
Gaussian function sources is taken as the form
\begin{equation}
\label{mat:8}ds^2=-h(r)dt^2+ h(r)^{-1}dr^2+r^2 d\Omega^2,
\end{equation}
where $h(r)=\left(1-\frac{2M_\theta}{r}\right)$, and
$d\Omega^2=d\vartheta^2+\sin^2\vartheta d\varphi^2$ is the line
element on the 2-dimensional unit sphere. The smeared mass
distribution $M_{\theta}$ is given by
\begin{equation}
\label{mat:9}M_{\theta}=\int_0^r\rho_{\theta}(r)4\pi
r^2dr=M\Bigg[{\cal{E}}\Big(\frac{r}{2\sqrt{\theta}}\Big) -\frac{
r}{\sqrt{\pi\theta}}e^{-\frac{r^2}{4\theta}}\Bigg].
\end{equation}
${\cal{E}}(x)$ is the Gaussian error function defined as $
{\cal{E}}(x)\equiv \frac{2}{\sqrt{\pi}}\int_{0}^{x}e^{-t^2}dt$. In
the commutative limit, $\frac{r}{\sqrt{\theta}}\rightarrow\infty$,
the Gaussian error function tends to one and the other term will
exponentially be reduced to zero. Thus, we get
$M_{\theta}\rightarrow M$. By using the Killing equation
$\partial_\mu\xi_\nu+\partial_\nu\xi_\mu-2\Gamma^\lambda_{\mu\nu}\xi_\lambda=0$,
and the condition of static spherical symmetry
$\partial_0\xi_\mu=\partial_3\xi_\mu=0$, and also the infinity
condition $\xi_\mu\xi^\mu=-1$, the timelike Killing vector of the
noncommutative Schwarzschild black hole is found to be
\begin{equation}
\label{mat:10}\xi_\mu=\left(-h(r),\,0,\,0,\,0\right),
\end{equation}
that is equal to zero at the event horizon.

According to Eqs.~(\ref{mat:1})-(\ref{mat:3}), the acceleration on
the spherical holographic screen with radius $r$ is computed as
\begin{equation}
\label{mat:11}a^\mu=\left(0,\,2\pi T,\,0,\,0\right).
\end{equation}
The above equation denotes that the screen carries a temperature in
the following form (using Eq.~(\ref{mat:3})):
\begin{equation}
\label{mat:12}T={1\over {4\pi}} {{dh(r)}\over {dr}}=\frac{M}{2\pi
r^2}\Bigg[{\cal{E}}\Big(\frac{r}{2\sqrt{\theta}}\Big)
-\frac{r^3+2\theta
r}{2\sqrt{\pi\theta^3}}e^{-\frac{r^2}{4\theta}}\Bigg].
\end{equation}
It should be noted that the local temperature on the event horizon
is the same as the Hawking temperature, i.e., $T|_{r=r_{H}}=T_H$
\cite{nic1,nic2} (see also \cite{meh1}).

The energy on the screen can be written in terms of the smeared mass
distribution as
\begin{equation}
\label{mat:13}E=2\pi r^2T
=M_{\theta}-\frac{Mr^3}{2\sqrt{\pi\theta^3}}e^{-\frac{r^2}{4\theta}}.
\end{equation}
Finally, the modified Newtonian force law as the entropic force in
the presence of the noncommutative Schwarzschild black hole becomes
\begin{equation}
\label{mat:15}F=\sqrt{g^{\mu\nu}F_\mu F_\nu}=
\frac{mM_{\theta}}{r^2}-\frac{mMr}{2\sqrt{\pi\theta^3}}e^{-\frac{r^2}{4\theta}},
\end{equation}
where $F_\mu=\left(0,\,\frac{m}{2\sqrt{h(r)}}{{dh(r)}\over
{dr}},\,0,\,0\right)$. We note that, in the limit of $\theta$ going
to zero, one recovers the conventional results for the acceleration,
the local temperature, the energy, and the entropic force on the
screen, respectively, as follows \cite{liu}:
\begin{equation}
\label{mat:16}a^\mu=\left(0,\,\frac{M}{r^2},\,0,\,0\right),\qquad
T=\frac{M}{2\pi r^2},\qquad E=M,
\end{equation}
and
\begin{equation}
\label{mat:19}F=\frac{mM}{r^2}.
\end{equation}
Eq.~(\ref{mat:19}) is just the Newton force for the Schwarzschild
black hole.

Now, as an important remark, we will show that the essential aspects
of the noncommutativity approach are not specifically sensitive to
the Gaussian nature of the smearing used by the authors in
Ref.~\cite{nic1}. If we had chosen a different form for the smeared
mass distribution, the general properties would be directed to
wholly similar results to those above. For example, we consider a
Lorentzian distribution of particle-like gravitational source as
follows
\begin{equation}
\label{mat:20}\rho_{\theta'}(r)=\frac{M\sqrt{\theta'}}{\pi^2(r^2+\theta')^2}.
\end{equation}
Here the noncommutativity parameter, $\theta'$, is not exactly the
same as $\theta$. The smeared mass distribution $M_{\theta'}$ is now
given by
\begin{equation}
\label{mat:21}M_{\theta'}=\int_0^r\rho_{\theta'}(r)4\pi
r^2dr=\frac{2M}{\pi}\left[\tan^{-1}\bigg(\frac{r}{\sqrt{\theta'}}\bigg)-\frac{r\sqrt{\theta'}}{r^2+\theta'}\right].
\end{equation}
The Lorentzian smeared mass, $M_{\theta'}$, has the same confining
properties and is totally similar to the Gaussian smeared mass,
$M_{\theta}$, qualitatively. In the limit $\theta'\rightarrow0$, we
get $M_{\theta'}\rightarrow M$. The local temperature on the screen
immediately reads
\begin{equation}
\label{mat:22}T=\frac{M}{\pi^2r^2}\left[\tan^{-1}\bigg(\frac{r}{\sqrt{\theta'}}\bigg)-\frac{r\sqrt{\theta'}}{r^2+\theta'}
-\frac{2r^3\sqrt{\theta'}}{(r^2+\theta')^2}\right].
\end{equation}
Using Eq.~(\ref{mat:21}), we obtain
\begin{equation}
\label{mat:23}T=\frac{M_{\theta'}}{2\pi r^2}
-\frac{2Mr\sqrt{\theta'}}{\pi^2(r^2+\theta')^2}.
\end{equation}
Eventually, the energy and the entropic force on the screen are
given by, respectively:
\begin{equation}
\label{mat:24}E=2\pi r^2T
=M_{\theta'}-\frac{4Mr^3\sqrt{\theta'}}{\pi(r^2+\theta')^2},
\end{equation}
and
\begin{equation}
\label{mat:25}F=\frac{mM_{\theta'}}{
r^2}-\frac{4mMr\sqrt{\theta'}}{\pi(r^2+\theta')^2}.
\end{equation}
\begin{figure}[htp]
\begin{center}
\includegraphics{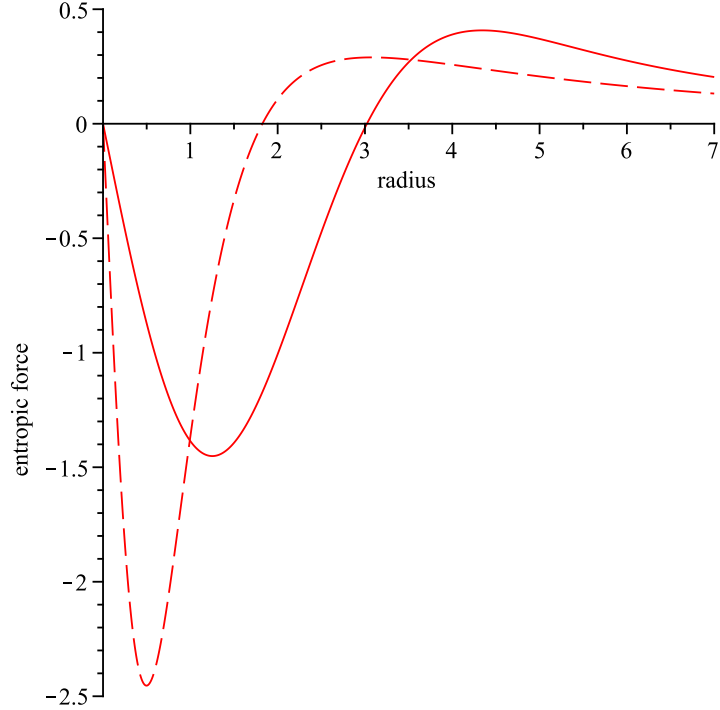}
\end{center}
\vspace{7.5 cm} \caption{\scriptsize {The entropic force as a
function of radius. The solid line, corresponds to the Gaussian
smearing and the dashed line, corresponds to the Lorentzian
smearing. We set $m=1.0$, and $M=10.0$. For plotting the figure thus
preserving the integrity of the outcomes, we set the value of the
noncommutativity parameter equal to unity in both Gaussian and
Lorentzian profiles; $\theta=\theta' = 1$.}} \label{fig:1}
\end{figure}

In order to facilitate comparison between Gaussian and Lorentzian
profiles, the results for the numerical solution of such
noncommutative entropic force as a function of radius in both
Gaussian and Lorentzian distributions (Eqs.~(\ref{mat:15}) and
(\ref{mat:25})) are depicted in Fig.~\ref{fig:1}. In both situations
(Gaussian and Lorentzian) a maximum $F$ happens at relatively small
$r$. As can be seen from the figure, within the noncommutative
geometry inspired model, the entropic force of the black hole grows
with the reduction of the radius of the spherical holographic screen
until it reaches to a maximum definite value (does not diverge at
all) and then falls down to zero at the minimal nonzero value of the
radius, $r_0$. The $r_0$ value is seen to be nearly similar in two
distributions. The minimal nonzero radius for the case of Gaussian
distribution is approximately equal to $3.0\sqrt{\theta}$, and in
the case of Lorentzian distribution one has
$r_0\approx1.8\sqrt{\theta'}$. Hence, most of the results that we
attained for the Gaussian case at least in asymptotic values of $r$,
remain valid if we choose the other case of probability distribution
of smeared matter.

As Fig.~\ref{fig:1} obviously shows the entropic force on the
holographic screen with radius $r_0$ is zero. This is a significant
result. The physical description of the $r_0$ is the radius of the
smallest holographic surface which can not be probed by a test
particle that is located within some distance from the source. We
have seen that the standard formulation of Newtonian gravity is
contravened at very short distance systems. In other words, when a
test mass $m$ has a distance $r_0$ from the source mass, it cannot
perceive any gravitational interactions. This phenomenon contravenes
the existence of gravitational interactions in an inert black hole
remnant \cite{adl}. Black hole remnants as essential entities are
widely accepted in quantum gravity literature when quantum
gravitational fluctuations emerge. For example, when generalized
uncertainty principle is taken into consideration, the total decay
of the black hole through radiation is prohibited, and we have
massive, but inert black hole remnants with only gravitational
interactions. Our approach clearly exhibits inert black hole
remnants with absolutely no gravitational interactions. On the other
hand, the equivalence principle of general relativity, which refers
to the equivalence of gravitational and inertial mass, is violated
because it is now possible to find a distinction between them. In
fact, the gravitational mass in the remnant size does not emit any
gravitational field, therefore it is experienced to be zero,
contrary to the inertial mass {\footnote{It should be noted that
there have been other schemes which violate the equivalence
principle such as the quantum phenomenon of neutrino oscillations
\cite{gas}, comparing Hawking radiation to Unruh radiation
\cite{dug}, and an examination of entropic picture of Newton's
second law for the case of circular motion \cite{dun} (somewhat more
related to the present work).}}. In spite the fact that it is
feasible to suppose both entropic gravity and noncommutative
geometry are true but in the case of short distances one expects a
contravention of the equivalence principle. In other words, it is
possible that when one comes close to the $r_0$ one deviates from
general relativity. Therefore, it is predictable that the
equivalence principle is violated at some small scale (maybe the
Planck scale) due to the combination of entropic gravity and
noncommutative geometry. Indeed there is a no-go theorem for
noncommutative Schwarzschild spacetime in Verlinde's approach to
entropic force. Otherwise, as mentioned one would violate the
equivalence principle.

In the case of $r<r_0$, one encounters some unusual dynamical
features leading to negative entropic force, i.e. gravitational
repulsive force (an exotic phenomenon, e.g. at the Planckian
regime). This means that if \,$r$ is extremely small, as the mass
$m$ approaches the screen, the decrease in screen entropy will
generate a repulsive force. However, we really should not trust the
details of our modeling when $r<r_0$. Most of the distinctions
between the Gaussian and Lorentzian profiles appear just in this
extreme region. In the region that noncommutativity effects begin to
be sensed exactly, the detailed nature of the sharpened mass
distribution is not indeed being inspected. As a matter of fact it
is possible for values $r\leq r_0$. But, it is important to
demonstrate that according to the original work proposed by Nicolini
{\it et al} \cite{nic1} (for a review see \cite{nic2}), concerning
the thermodynamics of the noncommutative inspired black holes, for
$M<M_0$ there is no solution for $g_{00}(r_H) = 0$ and no horizon
occurs, where $M_0$ is the minimal mass corresponding to an extremal
black hole with one degenerate horizon in $r_0$. So, there should be
a cut-off in the entropic force graph at some finite $r$, namely
$r_0$. In fact, if $r<r_0$ or the original mass is less than the
minimal mass $M_0$, there cannot be a black hole. Accordingly for
$r<r_0$ we cannot speak of an event horizon and no temperature can
be defined, so the final zero temperature configuration can be
considered a black hole remnant. This means that when the black hole
reaches the extremal configuration with a minimal mass, the
temperature is zero and the Hawking emission abruptly stops.

\begin{figure}[htp]
\begin{center}
\includegraphics{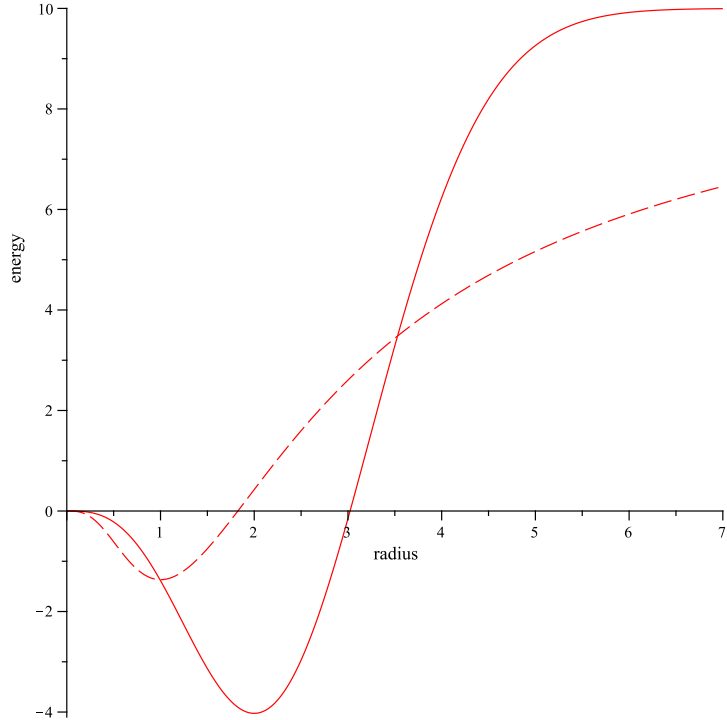}
\end{center}
\vspace{9.3 cm} \caption{\scriptsize {Energy versus radius. The
solid line, corresponds to the Gaussian smearing and the dashed
line, corresponds to the Lorentzian smearing. Other conditions for
plotting the figure are similar to Fig.~\ref{fig:1}. We set
$M=10.0$. For plotting the figure we set $\theta=\theta' = 1$. As
can be seen from the figure, the energy on a holographic screen with
radius $r_0$ is zero. In the case of $r<r_0$, the unusual feature,
i.e. $E<0$, is again evident in both situations.}} \label{fig:2}
\end{figure}

To clarify the issue even more we show the numerical evaluations of
Eqs.~(\ref{mat:13}) and (\ref{mat:24}) (the energy versus the radius
in both Gaussian and Lorentzian distributions) in Fig.~\ref{fig:2}.
This figure clearly exhibits that in the limit $r\gg\sqrt{\theta}$,
the energy on the screen is constant. The appearance of the black
hole remnant can also be seen as we approach regions with smaller
screen radii; this corresponds to Fig.~\ref{fig:1}. Hence, there is
also a same reason for the energy case as mentioned above for the
entropic force. Furthermore, the case of $E<0$ is nonphysical and a
finite cut-off is reasonable, therefore one can make the requirement
that $E\geq0$. Thus, if one considers the screen radius to be less
than the radius of the smallest holographic surface at the Planckian
regime, then one encounters some unusual dynamical features leading
to negative entropic force and negative energy. In a recent paper
\cite{meh2}, we have reported some results about exotic
thermodynamical treatment for Planck size black hole evaporation,
e.g. negative temperature, negative entropy, anomalous specific
heat, and etc., which may reflect the need for a fractal nature of
spacetime at very short distances. Theories such as $E$-infinity
\cite{nas} and scale relativity \cite{not} which are founded upon
fractal structure of spacetime may provide a suitable framework to
treat thermodynamics of these very short distance systems. On the
other hand, in quantum gravity regime, not only the geometry
containing a black hole is not truly static but also is surely
dynamic because black holes are accurately considered as highly
excited states. In other words, application of ordinary
thermodynamics to situations such as Planck scale black hole seems
to be impossible. Due to non-extensive and non-additive nature of
such systems, one should apply non-extensive formalism such as
Tsallis thermodynamics \cite{Tsa}. It seems that applying the
Tsallis thermodynamics into Verlinde's derivation is a fascinating
open problem {\footnote{Note that Ref.~\cite{net} is the first
article to apply Tsallis thermodynamics in the Verlinde
formalism.}}.

Finally, it should be noted that the value of $\sqrt{\theta}$ is
directly associated with the noncommutative scale and is assuredly
proportional to it. Nevertheless, the detailed nature of this
correlation is not clearly described and would need a more exact
framework to set it up. It is solely adequate to bear in mind that
$\sqrt{\theta}$ is proportional to the length scale or inverse mass
scale related to the noncommutative effects. The smallness of the
scale would indicate that noncommutativity influences can be
conceived just in excessive energy phenomena. In a general string
theory context one could suppose that $\sqrt{\theta}$ would
naturally not be far from the 4-dimensional Planck scale, $L_{Pl}$.
Most of the phenomenological investigations on noncommutative
approaches have suggested that we live in a 4-dimensional spacetime
and that the noncommutative energy scale is about $1-10$ TeV
\cite{hin}, accessible to colliders. Since, the minimal observable
length is not precisely defined through deduction; therefore the
scale is generally postulated as smaller than the typical scale of
the standard model of particle physics, i.e. only less than $
10^{-16} cm$. From another point of view, the fundamental Planck
scale in models with extradimensions \cite{ant} can be neighboring
current particle physics experiments \cite{ade}, and it may be
achieved in a TeV regime. Moreover, due to the appearance of
possible extra spatial dimensions in the TeV range, the abundant
creation of TeV-scale black holes at colliders prove feasible
\cite{bank}. Therefore it is acceptable to discuss that the
properties of such TeV-scale black holes may be affected by
noncommutativity influences, which are produced at a comparable
scale. If the noncommutative energy scale is actually of order of
the TeV range, then a direct probing of noncommutative physics may
be possible for instance at the LHC, ILC, CLIC or some other future
particle colliders.\\

\section{\label{sec:4}Summary and Conclusion}
In this article, we have discussed some aspects of Verlinde's
proposal in the presence of noncommutative Schwarzschild black hole
based on Gaussian-smeared mass and Lorentzian-smeared mass
distributions. The important difference between the two
distributions is apprehended to occur around $r_0$, where there
exists mostly the reactiveness to noncommutative effects and the
precise form of the matter profile. Nevertheless, both mass
distributions prepare the same quantitative aspects for the entropic
force. Among these aspects, it is worth of emphasizing: \textbf{(i)}
the fact that the gravitational and the inertial masses seem to be
separate contravenes the equivalence principle of general
relativity. However, one can come to this conclusion that
gravitational and inertial masses are not necessarily distinctive,
but the entire noncommutativity approach in gravity may be
incorrect. In fact, there is a no-go theorem for noncommutative
geometry inspired Schwarzschild black hole in Verlinde's scenario to
entropic force. Otherwise, one would contravene the equivalence
principle. \textbf{(ii)} at Planck scales either our analysis is not
the suitable one, or non-extensive statistics should be applied. In
fact, the comparison of the results obtained by the Gaussian profile
with the results that we obtained by the Lorentzian profile, leads
to two important conclusions: either we really cannot place a total
trust in the noncommutative effects with the Gaussian, Lorentzian
and some other cases of the smeared mass distribution at the regions
with the order of Planck length, or we actually should have some
misgivings in the results of standard thermodynamics at quantum
gravity level which the origin of this proposition may imaginably be
a result of the fractal nature of spacetime at very short distances.
These ambiguities seem to be good factors for a possible
experimental verification of Newton's law at very short distances in
the future. Indeed, at present we do not know which of these
decisions are reliable. However, we can always be
enthusiastic about the possibility of an experimental breakthrough.\\

\end{document}